\documentclass[structabstract]{aa}
\usepackage{txfonts}
\usepackage{epsfig,graphicx,mathptm}
\usepackage{times}
\usepackage{natbib}

\newcommand{\lcdm}{$\Lambda$CDM}
\newcommand{\omegam}{$\Omega_{\rm m}$}

\newcommand{\xmm}{{\it XMM-Newton}}
\newcommand{\chandra}{{\it Chandra}}
\newcommand{\ngal}{$N_{\rm gal}^{R200}$}
\newcommand{\mum}{$\mu$m}
\newcommand{\lsun}{$L_\odot$}
\newcommand{\iras}{IRAS}
\newcommand{\lir}{$L_{\rm IR}$}

\def\aap{A\&A}
\def\apj{ApJ}

\def\apjl{ApJ}
\def\mnras{MNRAS}
\def\araa{ARA\&A}
\def\aj{AJ}

\def\apjs{ApJS}
\def\pasp{PASP}

\begin{document}

\title{Infrared properties of the SDSS-maxBCG galaxy clusters}

\titlerunning{IR properties of the SDSS-maxBCG clusters}

\author{M. Roncarelli\inst{1},
        E. Pointecouteau\inst{1},
        M. Giard\inst{1},
        L. Montier\inst{1} and 
        R. Pello
        \inst{2} }

\authorrunning{M. Roncarelli et al.}

\institute{Centre d'Etude Spatiale des Rayonnements, CNRS/Universit\'e de 
           Toulouse, 9 avenue du Colonel Roche, BP44346, 31028 Toulouse 
           Cedex 04, France. \\
           \email{mauro.roncarelli@cesr.fr}
           \and
           Laboratoire d'Astrophysique de Toulouse-Tarbes, Universit\'e de 
           Toulouse, CNRS, 14 Av. Edouard Belin, 31400 Toulouse, France. \\
}

\date{Accepted ???. Received ???; in original ???}

\abstract{The physics of galaxy clusters has proven to be influenced by 
          several processes connected with their galactic component which 
          pollutes the intracluster medium (ICM) with metals, stars and dust. 
          However, it is not 
          clear whether the presence of diffuse dust can 
          play a role in clusters physics since a characterisation of  
          the infrared (IR) properties of galaxy clusters is very challenging 
          and yet to be completely achieved.}
         {In our study we focus on the recent work of Giard et al. (2008) 
          who performed a stacking analysis of the IRAS data in the direction of several 
          thousands of galaxy clusters, providing a statistical characterisation 
          of their IR luminosity and redshift evolution. We model the IR 
          properties of the galactic population of the SDSS-maxBCG 
          clusters ($0.1<z<0.3$) in order to check if it accounts for 
          the entire observed signal 
          and to constrain the possible presence 
          of other components, like dust in the ICM.}
         {Starting from the optical properties of the galaxies of the 
          SDSS-maxBCG clusters, we estimate their emission in the 60 and 
          100 \mum\ IRAS bands making use of modeled spectral energy distributions of 
          different 
          spectral types (E/S0, Sa, Sb, Sc and starburst). We also consider 
          the evolution of the galactic population/luminosity with redshift.}
         {The total galactic emission, which is dominated by the contribution 
          of star-forming late-type galaxies, is consistent 
          with the observed signal. In fact, our galactic emission 
          model slightly overestimates the observed fluxes, with the 
          excess being concentrated in low-redshift clusters 
          ($z \lesssim 0.17$).}
         {Our results indicate that, if present, the IR emission from 
          intracluster dust must be very small compared to the one associated 
          to the galaxy members. This translates into an upper limit on the 
          dust-to-gas mass ratio in the ICM of 
          $Z_{\rm d} \lesssim 5 \times 10^{-5}$. The excess in 
          luminosity obtained at low redshift constitutes an indication 
          that the cluster environment is driving a process of star-formation 
          quenching in its galaxy members.}

\keywords{cosmology: large scale structure of Universe -- 
          galaxies: clusters: general -- 
          intergalactic medium -- 
          infrared: galaxies}

\maketitle


\section{Introduction} \label{sec:intro}

Clusters of galaxies form in correspondence of the peaks of the 
primordial matter density field as a result of the 
gravitational collapse of both dark matter and baryons. In the 
framework of the standard \lcdm\ cosmological model and the 
hierarchical clustering of the large scale structure 
formation they constitute both the most recent and the 
largest virialised objects of the Universe. 

Nowadays, it is clear that gravitation is not the only process that
influences the physics of the intracluster medium (ICM, hereafter). 
The electrons
of the ionized plasma emit via free-free interaction with the protons,
making clusters bright X-ray sources and allowing the gas to cool
efficiently, particularly in the central regions. During the last
decade, the observations of the \xmm\ and \chandra\ X-ray satellites
highlighted the presence of several interaction mechanisms between 
the galactic component and the ICM, showing that the 
evolution of the two is intimately tied. For
instance, the accretion of cold gas onto the brightest cluster
galaxies (BCGs) at the cluster center is strongly suspected to fuel
the super-massive blackholes they host. This process is able to 
trigger 
star-formation within the BCGs \citep{odea08,pipino09} and power 
episodic violent outbursts of their central active galactic nuclei
(AGNs), whose energy injection into the ICM prevents the overcooling
of the gas \citep{fabian06,mcnamara07}. AGNs feedback and other 
non-gravitational processes, such as supernovae (SNe) powered 
galactic
winds \citep{kapferer06,sijacki06,schindler08}, preheating 
of the ICM \citep[see e.g.][]{fang08}, together with gravitational ones (i.e.
galaxy-galaxy interactions, ram-pressure stripping, galaxy mergers)
induce energy and matter exchanges between the galactic medium and the
ICM. To date these complex physical processes and their impact on the
statistical cluster properties, thus on our understanding of
structure formation \citep{voit05} and use of the
cluster population in cosmological studies
\citep{mantz08,vikhlinin09,mantz09}, are under scrutinous 
observational and
theoretical investigations \citep[see, e.g.,][and references
therein]{arnaud05,borgani08a}.

The presence of heavy elements in the ICM is the most
  direct evidence and consequence of the ejection of galactic material.  
  Within clusters their abundance has been widely
  measured making use of X-ray observations \citep[see the reviews
    by][]{sarazin88, arnaud05,werner08} with a typical abundance of
  0.3 $ Z_\odot$.
Stars and SNe constitute the most efficient way to produce and 
disperse metals. Heavy
elements in the ICM originate from different processes: early enrichment
\citep{aguirre07}, continuous injection from galaxy members and
in situ production by intra-cluster stars \citep[sources of the
intra-cluster light, see][]{arnaboldi04,krick07,murante07,conroy07,dolag09}.
The aforementioned processes do not discriminate between the natures 
of the ejected galactic material, therefore these enrichments 
are unambiguously 
linked also to the ejection of neutral gas and dust in the ICM.

Observations indicate that in galactic environments 
dust is a minor component, with dust-to-gas ratio 
$M_{\rm dust}/M_{\rm gas} \approx 0.01$ 
\citep{mathis77} and 
it could be as low as $M_{\rm dust}/M_{\rm gas} = 10^{-5}-10^{-4}$ 
in the ICM \citep{popescu00,aguirre01}. Nonetheless, dust particles 
have lifetimes long enough to be heated by collisions with the 
hot electrons and re-emit at the infrared (IR) wavelengths. 
In this 
way they can constitute an additional cooling agent 
of the gas \citep{montier04} which might play an important role in ICM 
physics, as seen with the implementation of dust cooling 
in hydrodynamical simulations \citep{dasilva09}.

However obtaining observational constraints on the possible 
IR signal coming from intracluster dust is a very challenging issue, since 
the average sky fluctuations caused by background galaxies 
and galactic cirrus clouds are comparable or even higher 
than the overall flux coming from a single cluster, which 
is anyway expected to be dominated by the dust emission from 
star-forming galaxies. In fact, nowadays the only claimed (and 
still controversial) detection of this 
IR emission comes from the studies on the Coma cluster of 
\cite{stickel98,stickel02}, who measured a diffuse dust mass of 
$M_{\rm dust} \approx 10^7 M_\odot$, while other attempts resulted 
in non detections \citep[see, e.g.,][]{bai07}.

If studies on single objects appear very problematic, the 
low signal-to-noise problem can be overcome by adopting a 
statistical approach, taking advantage of the high number 
of clusters detected mainly with optical surveys \citep[see]
[for a review]{biviano08}. This concept was first used at IR 
wavelengths by \cite{montier05} who performed a 
stacking analysis of the {\it Infra Red Astronomical Satellite} 
(\iras) survey in the direction of 
more than 11~000 known galaxy 
clusters and groups and obtained a statistical 
detection of the overall clusters signal at 12, 25, 60 and 100 
\mum. Starting from this result, \cite{giard08} characterised 
the IR luminosity evolution of the stacked sample, analysing 
also its correlation with their X-ray luminosities. These works 
constitute a first important result on the global IR properties 
of galaxy clusters. However, since the clusters IR emission 
could be dominated by star-forming galaxies, it 
becomes crucial to disentangle the galactic signal from a 
possible diffuse component in order to quantify its 
implications on cluster physics. On the other side, 
understanding the IR properties of the cluster galaxies could 
provide hints on the environmental effects over their star-formation, 
which are nowadays objects of studies at different wavelengths, 
from the radio to the ultraviolet \citep[UV, see][for a review]{boselli06a}.

In this work we attempt to reconstruct the observed stacked IR 
emission measured in the direction of galaxy clusters in the IRAS 
whole sky survey. We model in details the 
contribution of cluster galaxies to the total IR emission in order 
to understand whether this contribution is sufficient to explain the 
fluxes and luminosities derived by \cite{montier05} and 
\cite{giard08} or if, on the contrary, there is an indication of 
a non-galactic component, possibly associated to intracluster dust. 
Since for our modelisation we make use of the known 
spectral properties of galaxies as observed mainly in the field, the 
comparison between our results and observed data can 
be useful also to highlight possible differences between 
cluster and field galaxies.

This paper is organised as follows. In the next 
Section we briefly present the statistical detection of
the clusters IR emission of \cite{montier05} and \cite{giard08} and we
focus in particular on the subsample of the SDSS-maxBCG catalogue. In
Section~\ref{sec:model} we describe the details of our modelisation
for the IR luminosity of cluster galaxies and discuss the possible 
presence of other galactic components in Section~\ref{sec:other}. 
In Section~\ref{sec:flux}
we apply our model to estimate the \iras\ fluxes, then our
results are discussed in Section~\ref{sec:results}. Finally, we
present our conclusions in Section~\ref{sec:concl}.

Throughout this paper we assume a flat \lcdm\ cosmological model 
with $(\Omega_\Lambda,\Omega_{\rm m},h)=(0.7,0.3,0.7)$.


\section{The statistical IR emission of clusters} \label{sec:stack}

Working on a list of 11~507 groups and clusters
selected from 14 publicly available catalogues,\footnote{The list was
build making use of the SIMBAD database, operated at CDS,
Strasbourg, France (http://simbad.u-strasbg.fr/simbad/) and
NASA/IPAC Extragalactic Database (NED), operated by the
JPL/Caltech (http://nedwww.ipac.caltech.edu), among which the
ABELL catalogue \citep{abell89} and the {\it ``Northern Sky
Optical Cluster Survey''} \citep[NSC,][]{gal03}.}
\citet{montier05} performed a clear statistical detection of
the IR flux in the direction of galaxy clusters by 
stacking their corresponding fields (within 10' from the cluster center) 
in the \iras\ all sky survey. After dealing
  carefully with the point sources contamination, foreground/background
  subtraction and other various systematic effects, they measured
  stacked fluxes at 60 and 100~$\mu$m with signal-to-noise ratio of 57
  and 43 respectively. Over the four IRAS wavelengths this emission
  proves to be consistent with the spectral signature of galactic 
  IR emission.

More recently, on the basis of \cite{montier05} results,
\cite{giard08} performed a statistical analysis in redshift,
presenting for the first time the evolution of the IR luminosity of
galaxy clusters. In addition to \cite{montier05} original list of 
11~507 clusters,
\cite{giard08} backed-up their analysis on two 
{\it standalone} catalogues -- i.e.
the {\it ``Northern Sky Optical Cluster Survey''}
\citep[NSC,][]{gal03}, and the SDSS-maxBCG catalogue \citep{koester07}
-- for which a richness information was available.
They extended their IRAS stacking analysis to the {\it Rosat All Sky
  Survey} \citep[RASS,][]{voges92} in order to compare the evolution in
redshifts of the IR and X-ray luminosity.  They showed that the
stacked IR luminosities are on average 20 times higher than the X-ray
luminosities. They also found that the IR luminosity is evolving
rapidly as $(1+z)^5$ in the $0.1 < z < 1$ interval. \cite{giard08} 
also made use
of the richness information (i.e. correlated to the halo occupation
number), contained in the SDSS-maxBCG and NSC catalogues, to constrain
the dependence of the IR luminosities with cluster richness. They
derived a correlation following $L_{\rm IR} \propto (N_{\rm
  gal}^{R200})^{0.8\pm0.2} $.

In order to understand the aforementioned results at
60 and 100 \mum, in this work we focus on the modelisation of
the IR luminosity and fluxes of galaxy clusters. The IR spectrum 
of galaxies, which is dominated by dust emission
\citep{lagache05,soifer08}, is expected to make a major contribution 
to the IR emission detected in the direction of clusters \citep{giard08}. 
Therefore, a careful modeling of the IR emission of member galaxies
is needed in order to understand the global properties of clusters in
the IR, and in particular to disentangle between the relative contribution
of galaxies and intracluster dust.  Given the
inhomogeneity of the main dataset used by \cite{montier05} and 
\cite{giard08} (i.e. the 11~507 groups and clusters) in terms of cluster 
size and detection
method, it becomes very difficult to define a global selection
function of such sample, that would be necessary in order to 
have a suitable starting base for our
work. In order to overcome these difficulties we decided to base our 
modelisation on a restricted and well defined cluster sample by
considering a single catalogue. For the purpose of our calculations
we chose the SDSS-maxBCG catalogue.

\subsection{The SDSS-maxBCG catalogue} \label{ssec:maxbcg}

Since we want to reconstruct the total IR luminosity and flux due to 
cluster galaxies, we require an information on the halo occupation number 
(i.e. the number of galaxies within the cluster potential well). In this 
view the SDSS-maxBCG catalogue \citep{koester07} is the most fitted to 
our needs.  
Indeed, since the SDSS-maxBCG catalogue is created by
  identifying overdensities in the galaxy distribution, it contains
  information about the richness of the cluster itself that we can use
  directly without having to deal with the uncertainties connected
  with the $N-M$ and $L_X-M$ relations, that would be necessary, for
  instance, in case of X-ray selected clusters. Moreover, the number
  of objects contained in the SDSS-maxBCG catalogue is the highest
  among all the others (i.e. 13~807 groups and clusters), thus providing
  a good statistics which is needed to assess the robustness of our
  results.  Finally, \cite{koester07} provide also
  the total luminosity in the $r$ and $i$-band of the identified
  cluster members which also have a well defined spectral type 
  (E/S0 galaxies). We will make use of this information 
  to obtain an estimate of their flux in the \iras\ bands (see 
  Sect.~\ref{sec:model}).

The SDSS-maxBCG catalogue has been obtained by analysing the 
clustering properties of more than 500~000 SDSS galaxies in 
an area of $\sim$7~500 deg$^2$ in the redshift range 
$0.1<z<0.3$. For each cluster the catalogue contains a 
photometric redshift and an indication of the cluster 
richness, \ngal. This quantity corresponds to the 
number of early-type galaxy members at a distance lower 
than $R_{200}$ from the central BCG. The adopted definition 
of $R_{200}$ is the radius at which the deprojected galactic 
density is 200 \omegam$^{-1}$ times the average galactic 
density on the large scale: in the approximation of 
galaxies following the matter distribution, this is equivalent 
to the usual cosmological definition of $R_{200}$ as the 
radius enclosing a density 200 times larger than the critical 
density of the Universe \citep[see the discussion in][for details]
{hansen05}.

It is important to note that \ngal\ does \emph{not} represent the 
total cluster richness because it does not include spiral 
galaxies: in fact, the cluster members candidates have been 
selected for having colors matching the E/S0 ridgeline 
\citep{bower92} and $M_r<-16$. We will explain in the detail 
our definition of the cluster richness in Section~\ref{ssec:late}. 

The SDSS-maxBCG catalogue includes 13~823 groups and clusters with 
\ngal$\geq 10$ and it is widely dominated by the presence of 
groups and small clusters (\ngal$\approx 10-15$). As a reference, 
when computing their corresponding $M_{200}$ with the relation of 
\cite{rykoff08}, the median of the mass distribution is $4.5 \times 
10^{13} \, h^{-1} M_\odot$.
In this work, we adopted the trimming done by \cite{giard08} on 
the SDSS-maxBCG catalogue of  objects  located in regions of the 
sky not covered by the IRAS data, with noisy images, or whose 
fields contain strong IR sources. This selection process 
ended up in a sample of 7~476 clusters, corresponding to a total 
of 121~318 E/S0 galaxy members, that will be the object of our 
analysis.


\section{Modeling galaxy luminosities} \label{sec:model}

The main source of the IR radiation of galaxy clusters is expected 
to be the thermal emission due to the dust reprocessing the 
UV photons emitted by stars inside cluster galaxies. Since we are 
basing our work on galaxies observed in the optical band, we need 
to construct a valid model that is able to connect their galactic 
$r$ band emission 
to the corresponding dust emission in the IR. In particular, 
we will focus on the 60 \mum\ and 100 \mum\ \iras\ bands fluxes, in 
order to compare our results with the most relevant 
measurements of \cite{giard08}.
The characteristics of the galactic IR emission 
depend on the amount of dust present in the galaxies and are 
strictly connected with their star-formation rate (SFR) and, 
therefore, their spectral type.

\begin{figure}
\includegraphics[width=0.50\textwidth]{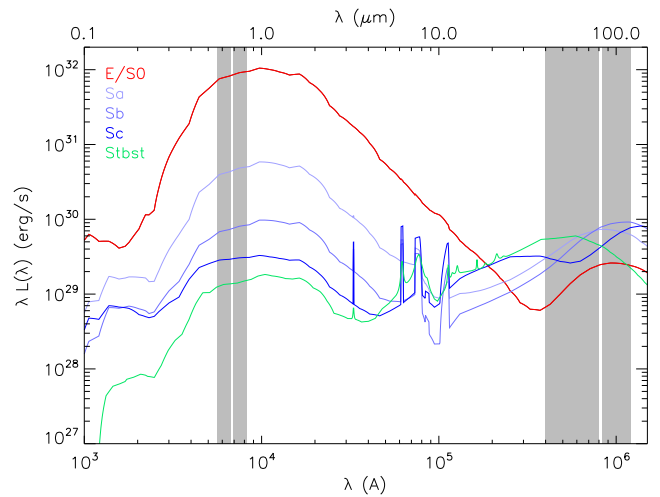}
\caption{SEDs as a function of wavelength for the 5
  \textsc{grasil} \protect \citep{silva98} templates used in this work. 
  The red line shows our reference
  model for E/S0 galaxies, the three blue lines represent normal
  spiral galaxies Sa, Sb, Sc (light blue, average, dark blue,
  respectively) and the green line represents our starburst galaxy
  model, corresponding to M82. All of the templates have been 
  arbitrarily normalized to the same integrated luminosity 
  of $10^{30}$ erg/s in the band 8-1000 \mum\
  (i.e. IR luminosity, \lir). The four shaded
  regions identify the optical $r$ and $i$-band and the two
  \iras\ bands of 60 \mum\ and 100 \mum, from left to right.  }
\label{fig:seds}
\end{figure}

Spectral energy distribution (SED) templates obtained with the spectral
synthesis code \textsc{grasil} \citep{silva98} were used in order to
represent the typical emission of galaxies in the local universe. 
\textsc{grasil} SEDs extend from UV to radio wavelengths, including
dust reprocessing and nebular line emission. We used four templates for normal 
galaxies, with Salpeter's initial mass function and age 13Gyr, namely an 
elliptical galaxy and
three different spiral galaxies (Sa, Sb and Sc). 
We also included the \textsc{grasil} model fit to
multi-wavelength observations of M82 as a semi-empirical template to represent
the typical SED of local starburst galaxies \citep{silva98}.  
For spiral galaxies, the SEDs correspond to a weighted average over the
different lines of sight, from face-on to edge-on, in order to statistically
account for the mean spatial orientations of cluster galaxies. 
The five templates are displayed in Fig.~\ref{fig:seds}.

The SED models described above were used to derive the expected IR
luminosities from observed fluxes and corresponding luminosities in the $r$
band. For a given template, the luminosity in the $r$ band is given by
\begin{equation}
L_r = \int{L(\lambda)T_r(\lambda)d\lambda} \ ,
\end{equation}
where $L(\lambda)$ is the luminosity per unit wavelength and $T_r(\lambda)$ is 
the SDSS $r$ filter transmission. For the purpose of this work, we 
define the IR luminosity following \cite{lefloch05}: 
\begin{equation}
L_{\rm IR} =
\int_{8\mu{\rm m}}^{1000\mu{\rm m}}{L(\lambda)d\lambda} \ .
\end{equation}
Therefore, for each template SED, the scaling ratio can be defined as follows
\begin{equation}
R_{{\rm IR},r} \equiv \frac{L_{\rm IR}}{L_r} \ .
\label{eq:f_ir}
\end{equation}

   Table~\ref{tab:gal_pop} summarizes the $R_{{\rm IR},r}$ values for the
different templates used in this paper. For a given $L_r$, the corresponding
\lir\ can be different by almost three orders of magnitude depending 
on the spectral type of the galaxy. The modelisation of the luminosity of 
early-type and late-type galaxy populations is explained in the details in
the next sections.

\begin{table}
\begin{center}
\caption{
Reference spectral types and galactic population models used in this work. 
}
\begin{tabular}{lrrrcr}
\hline
\hline
Spectral type     & \multicolumn{3}{c}{Population (\%)}  &  $u-r$ & $R_{{\rm IR},r}$  \\
                  & no ev. & ref. & cons. &        &                 \\
\hline
\it{-Early-type}  &      &      &      &             &        \\
BCG               &  4.3 &  4.0 &  4.2 &     --      &   0.12  \\
E/S0              & 65.4 & 61.5 & 63.3 &     --      &   0.12  \\
(Tot. early-type) &(69.7)&(65.5)&(67.5)&             &        \\
\\
\it{-Late-type}   &      &      &      &             &        \\
Sa                &  4.4 &  5.1 &  6.5 &  $>$2.2     &   2.36 \\
Sb                &  9.6 & 10.7 & 11.7 &  1.8--2.2   &  13.82 \\
Sc                & 15.8 & 18.2 & 14.1 &  1.0--1.8   &  33.80 \\
Starburst         &  0.5 &  0.5 &  0.2 &  $<$1.0     &  73.16 \\
(Tot. late-type)  &(30.3)&(34.5)&(32.5)&             &        \\
\hline
\hline
\label{tab:gal_pop}
\end{tabular}
\end{center}
The fraction (in percentile) of each type of galaxies in the three 
population models is shown: 
without redshift evolution, in our reference model and for the 
conservative scenario 
(second to fourth column, respectively). The fifth 
column shows the corresponding intrinsic $u-r$ color intervals 
used for the modelisation of the spirals (early-type galaxies are 
taken directly from the SDSS-maxBCG catalogue). The last 
column shows the values of the ratio 
$L_{\rm IR}/L_r$ for each template.
\end{table}

\subsection{Early-type galaxies} \label{ssec:early}

Early-type galaxies are characterized by an old stellar population and
a star-formation history which is essentially compatible with 
passive evolution: 
for this reason they are not expected to dominate the
emission in the \iras\ bands. In fact, for a fixed \lir, the 
amount of energy
emitted at wavelengths $\lambda \ga 40$ \mum\ is one order of
magnitude lower with respect to normal spirals (see
Fig.\ref{fig:seds}). However, it is known that in dense environments
elliptical galaxies largely dominate the galactic population, being
about 4 times more frequent than spirals \citep[see,
  e.g.,][]{dressler80,dressler97}. Moreover red galaxies are usually 
more massive
and more luminous in the $r$ band than blue ones. For these reasons
their total contribution on the clusters IR signal may be non-negligible.

As mentioned in Section~\ref{ssec:maxbcg}, \cite{koester07} identified
the number \ngal\ of E/S0 galaxies ($M_r<-16$) 
inside $R_{200}$ of each of the 7~476 
clusters of our sample. For each cluster, the authors provide the
luminosity, $k$-corrected at $z$=0.25, in the $r$ band.  of the BCG
and of the other E/S0 members as a whole. We corrected these
luminosities into rest-frame luminosities by using the LRG template of
\textsc{kcorrect} \citep{blanton07}: we refer to $L_{r}^{\rm BCG}$ and
$L_{r}^{{\rm memb}}$ for the luminosity of the BCG and of the cluster as a
whole, respectively, after this correction. Given the uniform 
properties of early-type galaxies, and the high number of
such objects in our sample, their typical SED should be well 
represented by the E/S0 template described above. In fact, although 
the IR signal of every single galaxy is lost in the stacking process, 
the E/S0 template still represents the average behavior of the 
early-type population of galaxies.

For every cluster we assign a $r$-band luminosity 
to all of its early-type members contained in the SDSS-maxBCG 
catalogue. For the BCG we use the value of 
$L_{r}^{\rm BCG}$, while for the other galaxies we use the 
average luminosity of the non-BCG E/S0 galaxies, determined 
as
\begin{equation}
L_{r}^{\rm avg} = 
\frac{L_{r}^{\rm memb}-L_{r}^{\rm BCG}}{N_{\rm gal}^{R200}-1} \ .
\label{eq:lr_others}
\end{equation}
From eq.~\ref{eq:f_ir}, we translated these $r$-band luminosities 
into IR luminosities: $L_{\rm IR} = 
R_{{\rm IR},r} \times L_r$, subsequently used to normalise the 
SED and further on to derive the fluxes in the 60 \mum\ and 100 
\mum\ \iras\ bands (see  Sect.~\ref{sec:flux}).

\subsection{Contribution of late-type galaxies} \label{ssec:late}

Although the majority of optically bright galaxies in clusters 
environment are
elliptical, and despite spiral galaxies in dense
environments tend to be quickly stripped of their gas and have 
their star-formation quenched, they are still expected to provide a 
dominant contribution to the IR emission, due to their higher
SFRs. Therefore, the contribution of late-type galaxies to the 
IR emission is crucial for our purposes and should be carefully 
estimated. Given the fact that  
late-type galaxies are not considered in the SDSS-maxBCG catalogue, 
their contribution was computed based on available modelings 
of galaxy populations, in particular the distribution of 
galaxies as a function of spectral type and luminosity. Since 
our sample goes from galaxy groups to rich clusters, we have
included the environment dependence of these variables. 

    We based our model on the known properties of galaxies in the local
universe as given by \cite{balogh04} who analysed the bimodal distribution
of galaxies in a local ($z<0.08$) sample of SDSS galaxies (DR1). These authors
performed a detailed study on the color and luminosity distribution of
galaxies as a function of their environment for both early and late-type
galaxies.  For each galaxy they define a density estimator
$\Sigma_5$ which represents the local projected galaxy density ($M_r<-20$) in
Mpc$^{-2}$ and they use it to classify the different environmental
regimes. Beside \ngal\ mentioned before (see Sect.~\ref{ssec:maxbcg}), the 
SDSS-maxBCG 
dataset contains also information about the number
of galaxies $N_{\rm gal}$ projected within a distance lower than 1
$h^{-1}$ Mpc from the BCG, with the magnitude limit of $M_r<-16$. 
Therefore for each cluster we can assume a
unique value of $\Sigma_5$ for all of its members by establishing a
relation with $N_{\rm gal}$ and, finally, with the properties of 
the population of late-type galaxies present.

We proceed as follows.

\begin{itemize}
\item[i)]First of all, we integrate the double gaussian 
distributions of red and blue galaxies 
\citep[Fig. 1 of][]{balogh04} obtaining a ratio of red and blue 
galaxies for the 5 values of $\Sigma_5$ considered: for our work, we 
only need the three density bins corresponding to dense environments (i.e. 
groups and clusters, $\Sigma_5 > 0.5$). In these bins  the resulting spiral
fractions are $f_{\rm spi}=$0.597, 0.478 and 0.286, in order of
increasing density.

\item[ii)] In order to use these values for our 
calculations we need, at first, to correct the value of $N_{\rm gal}$.
The galaxy samples of 
\cite{koester07}, in fact, has a magnitude limit 
$M_r < -16$ while the definition of $\Sigma_5$ includes only 
galaxies with $M_r < -20$: for this reason we use the red 
galaxies luminosity function (LF) given by \cite{baldry04} to 
calculate the number of expected galaxy members with 
$M_r < -20$. We obtained $<N_{\rm gal}^{-20}>=0.32 \, N_{\rm gal}$. 
As for \cite{balogh04}, the sample of \cite{baldry04} is constituted 
by local SDSS galaxies, with the difference that it contains no 
density distinction (e.g. field galaxies). Therefore, by 
using the LF of \cite{baldry04} for this 
calculation we neglect the effect of environmental properties; 
anyway, since red galaxies are mostly present in galaxy 
clusters, we do not expect this to change significantly our 
results.

\item[iii)] We are now able to obtain a first estimation 
of the density parameter associated to every cluster with 
the formula
\begin{equation}
\Sigma_{5,0} = \frac{<N_{\rm gal}^{-20}>}{1-f_{{\rm spi},0}} 
               \frac{h^2}{\pi (1 {\rm Mpc})^2} \ ,
\label{eq:sigma5}
\end{equation}
where $f_{{\rm spi},0}$ is the fraction of spiral galaxy 
(with $M_r < -20$) that for this calculation we fix arbitrarily 
to the initial value to $f_{{\rm spi},0}=0.3$.

\item[iv)] Using the value of $\Sigma_{5,0}$ we 
calculate a new estimate  $f_{{\rm spi},1}$ of the spiral 
fraction by interpolating between the values of 
$f_{\rm spi}$ obtained by \cite{balogh04}; this allows the 
calculation of a new value $\Sigma_{5,1}$ of the galaxy density 
with equation~\ref{eq:sigma5}. 

\item[v)] We repeat this procedure until $\Sigma_{5,i}$ and 
$f_{{\rm spi},i}$ converge to definite values and check that 
the result does not depend on the initial choice of 
$f_{{\rm spi},0}$, thus obtaining the expected number 
of spiral galaxies with $M_r < -20$
\begin{equation}
<N_{\rm spi}^{-20}> = <N_{\rm gal}^{-20}> \times 
\frac{f_{{\rm spi}}}{1-f_{{\rm spi}}} \times
\frac{N_{\rm gal}^{R200}}{N_{\rm gal}} \ ,
\end{equation}
where the factor $N_{\rm gal}^{R200}/N_{\rm gal}$ accounts for the 
ratio between the galaxies inside $R_{200}$ and $1 h^{-1}$ Mpc.

\item[vi)]Finally, we add the expected number of spiral galaxies in the
magnitude range $-20 < M_r < -18$ by referring again to the galaxy
distributions of \cite{balogh04}, thus obtaining for every cluster the
estimated value of the number of spiral members with
  $M_r < -18$, $<N_{\rm spi}>$. This value is then used as the
average of a poissonian distribution, that we assumed
  to assign a randomised number of spiral $N_{\rm spi}$ to each
  cluster, thus to determine the total cluster galaxies: $N_{200} =
N_{\rm gal}^{R200}+N_{\rm spi}$. 

\end{itemize}

\begin{figure}
\includegraphics[width=0.50\textwidth]{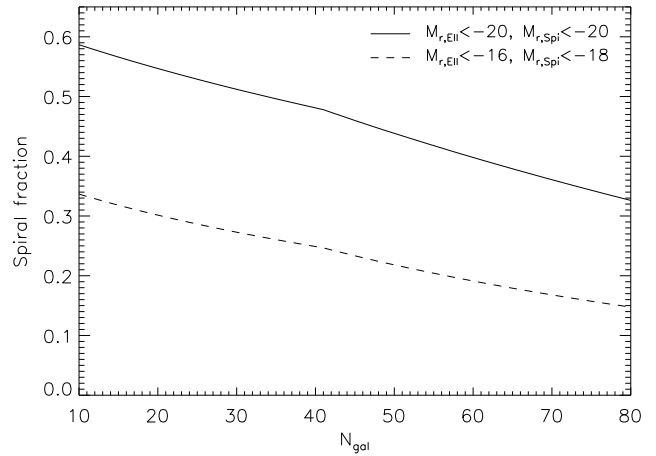}
\caption{Fraction of spiral galaxies within $R_{200}$
  (see Sect.~\ref{ssec:maxbcg} for the definition of $R_{200}$) as a
  function of $N_{\rm gal}$. The solid line represents the result
  obtained when considering only galaxies with $M_r < -20$ \protect 
  \citep[i.e. the magnitude limit used to calculate $\Sigma_5$ in][]
  {balogh04}. The dashed 
  line is the result obtained by
  considering all the galaxies included in our model: late-type
  galaxies with $M_r < -18$ and early-type galaxies with $M_r < -16$
  from the SDSS-maxBCG catalogue.   These
  functions have been calculated neglecting the evolution of the
  spiral fraction with redshift.}
\label{fig:spiral_frac}
\end{figure}

This formalism led us to derive an average fraction of
about 30\% spiral galaxies in our sample of SDSS-maxBCG clusters
(see Table~\ref{tab:gal_pop}). 
The relation between $N_{\rm gal}$ and the 
fraction of spiral galaxies inside the whole cluster is shown in Fig. 
\ref{fig:spiral_frac}. When considering galaxies with the 
same magnitude limit (dashed line), the spiral fraction 
can be higher than 0.5 for groups and small clusters, and then 
lessens with increasing $N_{\rm gal}$ down to $\sim$0.4 for 
$N_{\rm gal}$=70. When considering all the galaxies included in 
our model (i.e. $M_r<-16$ for early-type and $M_r < -18$ for late-type, 
solid line) the spiral fraction becomes lower by about 0.2.

The morphology-density study of \cite{balogh04} becomes 
useful also to assign a magnitude and a color to each 
spiral galaxy. For a given value of $\Sigma_5$ it is 
possible to construct a probability distribution 
$\mathcal{P}(M_r,u-r)$ for a late-type galaxy 
to have a given magnitude $M_r$ and a given intrinsic color $u-r$. 
The former value is used to determine the luminosity $L_r$ in 
the $r$-band as 
\begin{equation}
L_r = L_{0,r} 10^{-0.4M_r} \ ,
\label{eq:mag2lum}
\end{equation}
where $L_{0,r}$=2.15$\times$10$^{34}$ erg/s is the luminosity of 
an object with $M_r=0$, while the latter is used to assign the 
spectral type, according to the intervals shown in Table~\ref{tab:gal_pop}.
To define these intervals we calculated the $u-r$ colors of our 
\textsc{grasil} templates (2.40, 2.01 and 1.55 for the Sa, Sb and Sc, 
respectively) and defined the limits in order to associate to every 
galaxy the template that best approximates its value of $u-r$. 
We also checked the consistency of these numbers with the observed 
colors of local galaxies as reported by \cite{fukugita95} for the 
SDSS photometric system\footnote{Referring to their Table~3, they 
obtain $u-r$ colors of 2.26 for Sab and 1.68 for Sbc, thus close to 
our assumptions for the Sa/Sb and Sb/Sc limits, respectively.}.
For what concerns starburst galaxies, the limit of 
$u-r < 1.0$ has been taken directly from \cite{fukugita95} for irregular 
galaxies, since the $u-r$ of our M82 templates ($u-r=2.10$) is not 
representative of the starburst galaxies population. However, we will 
show that the choice of this limit has a negligible impact on our 
final results. Finally, the luminosity $L_{\rm IR}$ 
is obtained from Eq.~\ref{eq:f_ir},
where $R_{{\rm IR},r}$ is chosen according to the spectral 
type.

\subsection{Evolution with redshift of spiral galaxies} \label{ssec:z_evol}

  Despite being the SDSS-maxBCG a catalogue of relatively nearby
  groups and clusters ($0.1<z<0.3$), it is necessary to take into
  account the evolution with redshift of the late-type galaxies
  properties with respect to the results of \cite{balogh04}
 which are derived from a local (i.e. $z<0.08$) sample
of SDSS galaxies. For the purpose of our modelisation we consider the
sample of \cite{balogh04} as representative of the galaxy population
at $z_{\rm B}=0.04$, the central value of the interval,
and apply two independent evolution effects. \newline
\newline
{\bf Spiral fraction evolution} \newline
It is generally accepted that in a hierarchical structure formation 
scenario spiral galaxies tend to evolve towards S0 or ellipticals. 
This can happen either passively, with the consumption of the gas 
reservoir and its ejection with SNe explosions, or it can also 
be triggered by environmental interactions with other galaxies 
and with the ICM \citep[see][for a review]{boselli06a}: it is 
therefore expected that 
the fraction of late-type galaxies increases with redshift 
\citep[the so called Butcher-Oemler effect,][]{butcher84}. 
In fact, \cite{lagana09} find a change from about 0.1 to 0.3 
in the spiral fraction of 20 galaxy clusters in the 
interval $0<z<0.25$. Fitting the data on the fraction of 
star-forming galaxies of these clusters, they quantify its 
redshift evolution as: $\frac{df_{\rm spi}}{dz} = 1.3\pm 0.6$ 
(private communication). A similar result has been 
obtained by \cite{delucia07}, who find $\frac{df_{\rm spi}}{dz} 
\sim 1.1$ in the range $0.4<z<0.8$. Given the redshift interval 
of the SDSS-maxBCG catalogue, we take as a reference the value of 
\cite{lagana09}. Despite this sample contains rather massive 
objects ($k_{\rm B}T$ from $\sim3$ to $\sim10$ keV), it should provide 
a first-order representation of the spiral fraction evolution 
in the redshift range of the SDSS-maxBCG catalogue. We 
therefore correct the number of spiral galaxies obtained in 
Section~\ref{ssec:late} with the following formula:
\begin{equation}
<N_{\rm spi}(z)> = [1+1.3 \, (z-z_{\rm B})]<N_{\rm spi}>_{z=0} \ .
\end{equation}

Including this effect in our modelisation raises the global spiral 
fraction of our sample from 30\% to 35\%, as shown in 
Table~\ref{tab:gal_pop}.
Although this number may seem high for clusters \citep[see e.g.]
[for the Coma cluster]{bai06} we must consider that the SDSS-maxBCG 
catalogue is dominated by groups and small clusters, as mentioned 
in Section~\ref{ssec:maxbcg}. However, when applying our model to 
the case of MS 1054-03 as observed by \cite{bai07} 
($z=0.83$, $N_{200}=144$), we obtain $f_{\rm spi}=0.16$ which is 
consistent within a 1$\sigma$ limit with the value of $0.13 
\pm 0.03$ found by the authors.
\newline
\newline
{\bf Luminosity evolution} \newline The current hierarchical structure
formation scenario predicts that the peak of the star-formation is to
be placed at $z \approx 2-3$ and that at later epochs galaxies are
mainly consuming their reservoir of gas. For this reason, when
modeling the IR emission in the redshift range of the SDSS-maxBCG cluster
sample we must consider that higher SFRs are expected with respect to
the local galaxies. This effect has been widely observed and
quantified. We take as a reference the evolution of the IR-luminosity
observed by \cite{lefloch05}: when parameterising it in the form of
$L_{\rm IR} \propto (1+z)^{\alpha_L}$, they obtain $\alpha_L=3.2^
{+0.7}_{-0.2}$ over a sample of 2 635 objects identified at 24 \mum\ 
in the \chandra\ Deep Field South in the redshift range $0 < z \la 1$. 
This result has been
obtained by analysing field galaxies and this might introduce an
environment bias on our results. Anyway, since it is expected that
local star-forming galaxies inside clusters were the latest accreted, 
we can assume that this evolution scenario is valid also in cluster 
and group environments. Moreover, as shown in figure~3 of \cite{giard08},
this assumption is backed-up by the {\it Spitzer} observation of the
distant cluster MS1054-03 \citep{bai07}. Another bias may be introduced 
by the fact that these objects were identified directly in the mid-IR 
and not in the optical as for the galaxies of our sample. Therefore 
the galaxies of this sample are brighter than the ones of 
the SDSS-maxBCG, both in the IR and in the optical.
However, 
the total luminosity of the sample of \cite{lefloch05} is dominated by 
luminous (\lir$> 10^{11} L_\odot$) massive galaxies, which are the ones 
that provide most of the signal also in the \iras\ bands. Therefore, we 
assume that their result 
can be applied also to the SDSS-maxBCG cluster sample to describe its 
global IR luminosity evolution.
Consequently, we determine 
the IR luminosity $L_{\rm IR}$ as a function of redshift:
\begin{equation}
L_{\rm IR}(z) = 
L_{{\rm IR},0}(1+z-z_{\rm B})^{3.2} \ ,
\end{equation}
where $L_{{\rm IR},0}$ is the luminosity obtained 
by the magnitude $M_r$ as from eq.\ref{eq:mag2lum}.

We do not include any specific effect of spiral population 
evolution towards bluer types with increasing redshift. In fact, 
the fraction of the different spectral types of spirals in the 
redshift range we are interested in is not expected to differ 
significantly with respect to the local galaxies inside galaxy 
clusters \citep[see e.g.][]{delucia07}. Moreover the consequent 
effect of slightly increased global luminosity is anyway 
partially included when assuming the luminosity evolution 
with redshift aforementioned.

\subsection{Uncertainties on the model parameters}
\label{ssec:unc}

The choice of the values of the different parameters
introduces a degree of uncertainty in our modelisation, the most
important of which are connected with the choice of the redshift
evolution parameters and the $u-r$ color intervals used to define the
spiral types (see Table~\ref{tab:gal_pop}). In order to quantify these
uncertainties, we introduce two other sets of parameters to define a
conservative and an extreme scenario and we will use the values of the
fluxes estimated with these models to define the degree of confidence
of our results.  More precisely, for what concerns the redshift
evolution we adopt the $\pm 1 \sigma$ values for $\frac{df_{\rm spi}}
{dz}=(0.7,1.9)$ and $\alpha_L=(3.0,3.9)$, as measured by
\cite{lagana09} and \cite{lefloch05}, respectively. Moreover, we also
artificially add a shift in the $u-r$ color intervals of $\pm 0.1$ in
order to obtain a redder (conservative scenario) and a bluer (extreme)
spiral population. Anyway, these changes introduce only minor
modifications in the galaxy population: the difference in the 
global spiral fraction is of $\sim2$\%, while the only significant 
change is in
the fraction of starburst galaxies that goes from 0.2\% in the
conservative model to 1.1\% in the extreme one. On the whole, the 
difference in the final fluxes is mainly introduced by the modified  
luminosity evolution. The results on the galaxy
population for the conservative model are also shown in
Table~\ref{tab:gal_pop}.

\subsection{The IR luminosity function of the SDSS-maxBCG galaxies}
\label{ssec:lumfun}

\begin{figure}
\includegraphics[width=0.50\textwidth]{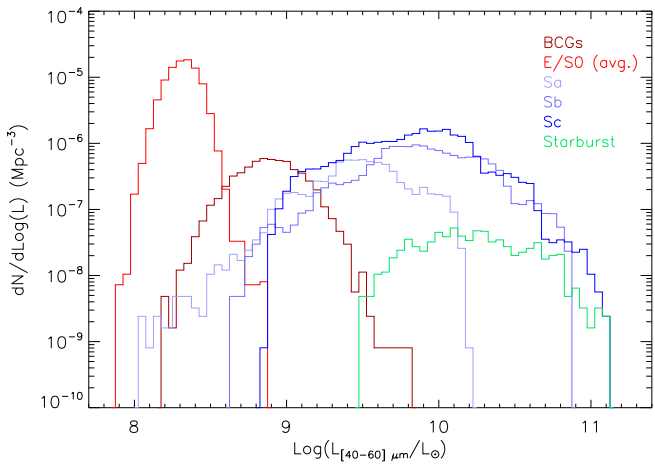}
\caption{Luminosity functions of each spectral type 
  implemented in our model in the \iras\ 60~\mum\ (i.e. top hat
  band-pass within [40,80]~\mum). Spectral types are color-coded
  as follow: (i) early-type galaxies as (dark-red) for the BCGs and
  (light-red) for the other E/S0s (we assigned to every galaxy the
  average luminosity of the E/S0 members of each cluster, as shown in
  eq.~\ref{eq:lr_others}); (ii) normal late-type galaxies, with 
  Sa, Sb, Sc going from (light-blue) to (dark-blue), respectively; 
  (iii) starburst galaxies as (green).}
\label{fig:lum_func}
\end{figure}

We show in Fig.~\ref{fig:lum_func} the luminosity function of 
the galaxies included in our reference model, for the different 
spectral types. The luminosity considered is given in the 
rest-frame 60 \mum\ \iras\ band.

The luminosities go from a minimum of $\sim10^8$\lsun\ for the 
fainter E/S0 galaxies up to a maximum of some $10^{11}$\lsun\ 
for the brightest spirals. 
Each type of spiral has a luminosity range of about 2 orders 
of magnitudes: this is a consequence of our method (see 
Sect.~\ref{ssec:late}), that held the spiral magnitudes within 
the interval $-23<M_r<-18$ which corresponds to a factor of 
$10^{0.4 \ \Delta M_r}=100$ in luminosity 
with the only possible modification introduced by the 
redshift evolution. Anyway, we don't expect this limitation 
to affect significantly our results. In fact, this limit 
range is wider than the one covered by the distribution of the BCGs, 
which are the only galaxies for which we have direct 
observational constraints. On the contrary the E/S0 curve 
appears narrower because we considered the average 
luminosity of the E/S0 galaxies inside each cluster, thus 
limiting the dispersion of the E/S0 galaxies to the 
dispersion of the cluster luminosities.

When considering the luminosity function in the 100 \mum\ 
\iras\ band the range and shape is similar for all the 
spectral types. Only starbursts are a factor $\sim$3 fainter 
in this band due to the drop in luminosity (see 
Fig.~\ref{fig:seds}) which is due to an average 
higher galactic dust temperature compared to non-starburst.

\section{Other galactic components}
\label{sec:other}
In this Section we try to put some constraints on the impact 
of three components of galactic origin which are not included 
in our modelisation of the IR emission of cluster galaxies: 
a missed population of faint galaxies, the possible presence of 
dust-embedded AGNs and the IR emission coming from 
heavily obscured star-forming galaxies. In all of these cases we 
conclude that their impact on our final results is probably 
very small, if not completely negligible.

\subsection{Faint galaxy population}
As mentioned in Sect.~\ref{sec:model}, our modelisation includes 
red galaxies with $M_r < -16$ (the limit of the SDSS-maxBCG 
catalogue) and blue galaxies with $M_r < -18$, thus 
neglecting the signal of fainter objects, that may not be 
identified by the SDSS observations although present inside the 
cluster. We try to obtain 
a rough estimation of the impact of these objects on the 
total emission by referring to the luminosity functions 
obtained by \cite{baldry04} (see their Fig. 7).

For what concerns early-type galaxies, the faint-end of the 
LF ($M^*=-21.49$, $\alpha=-0.83$ in the Schecter function 
parameterisation) has a negative slope, so the number of galaxies  
is expected to diminish at higher magnitudes. For this 
reason no significant impact can be associated to these 
objects.

On the other side, the faint-end of the blue galaxies LF 
($M^*=-20.60$, $\alpha=-1.35$) has a positive slope, so an 
increasing 
number of galaxies is expected with lower luminosities. 
For simplicity, we consider the assumption that all 
galaxies with $M_r < -18$ have been included in our 
modelisation, then we integrate the late-type LF 
$\phi_{\rm b}(M_r)$ of \cite{baldry04} splitting it in 
two at this magnitude limit. We obtain
\begin{equation}
\frac{L_{\rm faint}}{L_{\rm bright}} = 
\frac{\int_{-18}^\infty \phi_{\rm b}(M_r)L(M_r)dM}
     {\int_{-23}^{-18}\phi_{\rm b}(M_r)L(M_r)dM}
     = 0.13 \ ,
\label{eq:lfaint}
\end{equation}
where $L(M_r)$ is the luminosity as a function of the 
magnitude $M_r$, thus indicating that the emission of the 
faint galaxy population is marginal with respect to the 
bright one. 

Moreover, since the LF of \cite{baldry04} is 
obtained with the observation of field galaxies, it is 
reasonable to expect that in cluster environments the 
faint-end of the LF would be shallower, if not even 
negative, due to the processes of merging that affect 
particularly the smaller objects. Therefore the ratio 
of 0.13 obtained in eq.~\ref{eq:lfaint} can be safely 
considered an upper limit of the true value.
For these reasons, and considering the assumption of the 
connection between the luminosity in the optical and IR 
bands used throughout this work, we can conclude that even 
if we can not exclude the presence of the signal of 
a population of faint ($M_r > -18$) unresolved galaxies, 
we expect that the contribution of faint star-forming 
galaxies to be marginal.

\subsection{Dusty AGNs}
When estimating the total IR luminosities of the galaxies 
of our sample, we do not take into account what could be 
the contribution of AGN deeply embedded in dusty cocoons 
within cluster galaxies.

In fact, recent results indicate the existence of 
a population of heavily absorbed AGNs in the field \citep[see e.g.][]
{fiore09,lanzuisi09}: these objects are detectable at IR wavelengths.
Although optical observations indicate a very small fraction of
cluster galaxies with detected AGNs \citep[$\sim1$\%, see
  e.g.][]{dressler99}, some X-ray observations have detected an excess
of point sources associated to AGNs in cluster fields. For example,
\cite{martini06} observed with {\it Chandra} a sample of 8 low redshift
($z \lesssim 0.3$) clusters, finding $\sim$5\% of their galaxy members 
hosting an AGN, most of which are not detected with optical
surveys.
These AGNs are also present in E/S0 galaxies, and whilst
they are obscured in the optical (their emission being heavily
absorbed), they could be bright in the IR. Therefore, they could 
contribute to the total IR cluster emission.

It is difficult to quantify with precision the impact on the total IR
signal of these kinds of objects. We take as a reference the work done
by \cite{bai07}, who studied the IR properties of the galaxies of
MS1054--03 ($z=0.83$) with the {\it Spitzer}
satellite. They identified eight point sources with X-ray and
radio observations that could be associated with AGNs: anyway, only
three of these objects have an IR counterpart over the 144 IR-detected
cluster members, and only one of these is associated with a bright
star-forming object.  For these reasons the authors conclude that the
contamination from AGNs is negligible. Similar conclusions have been
drawn by \cite{bai06} in their study of the Coma cluster ($z=0.023$).

Even if they are based on single cluster studies, 
these results, which have been obtained both at 
redshift higher and lower than our cluster sample, 
indicate that the impact of AGN contamination in 
the observations of \cite{giard08} is probably 
very small.

\subsection{Heavily obscured star-forming galaxies}
Optical observations have highlighted the fact that some 
galaxy clusters host heavily obscured star-forming galaxies. 
In the framework of our modelisation, the presence of these 
objects would lead to an underestimate of their IR emission 
based on the optical one \citep[see][for a review]{metcalfe05}. 
In particular, IR observations on A1689 \citep{duc02}, 
J1888.16CL \citep{duc04}, 
CL0024+1654 \citep{coia05} and, more recently, on A1758 
\citep{haines09} have revealed galaxies with much higher SFRs 
with respect to what is expected from optical diagnostics (e.g. 
[O{\sc ii}]). However, all of these clusters show clear signs 
of recent dynamical activities, like major mergers and 
significant accretion of galaxies from the field, thus 
suggesting that these phenomena are responsible of the 
star-formation triggering. Since our cluster sample is at a 
relatively low redshift, we do not expect it to contain a high 
fraction of dynamically active haloes and, therefore, to be 
significantly affected by the presence of heavily obscured 
star-formation.


\section{Reconstructing the stacked IR flux} 
\label{sec:flux}

As mentioned before the main objective of this work is
to explain the origin of the IR emission observed in the direction
of galaxy clusters by \cite{montier05} and \cite{giard08}. Therefore, 
we want to compare
the flux expected out of our model from the galactic emission in the
60 \mum\ and 100 \mum\ \iras\ bands with the one measured by
\cite{giard08}. In order to do so, we need to compute for every
single cluster the expected flux taking into account the instrumental 
beam and the spectral band pass of the \iras\ satellite.

We assume that every galaxy is placed at the 
redshift $z$ of the cluster to which it belongs and we compute 
its flux $F_{\lambda_0}$\footnote{This quantity corresponds to a 
flux per unit frequency (e.g. Jy). Anyway we prefer to use the 
notation $F_\lambda$, rather than $F_\nu$, to refer directly to the 
fluxes $F_{60}$ and $F_{100}$ in the 60 and 100 \mum\ bands, 
respectively.} 
in a given band
\begin{equation}
F_{\lambda_0} = 
\frac{1}{4\pi d_{\rm L}^2 \Delta \nu_0}
\int L \left( \frac{\lambda}{1+z} \right)
f_{\lambda_0}(\lambda) d\lambda \ ,
\end{equation}
where $\Delta \nu_0$ is the frequency interval of the corresponding band, 
\begin{equation}
d_{\rm L}(z) = \frac{c(1+z)}{H_0}\int_0^z 
               \frac{dz'}{\sqrt{\Omega_{\rm m}(1+z')^3+\Omega_\Lambda}}
\end{equation}
is the luminosity distance of the cluster 
($H_0 \equiv 100\, h$ km s$^{-1}$ Mpc$^{-1}$),
$f_{\lambda_0}(\lambda)$ is the \iras\ spectral response function 
in a given band (i.e. $\lambda_0 =60, 100$ \mum) which is
convolved\footnote{For these calculations we used the 
tools of the \textsc{dustem} code, an updated version of the 
emission model of \cite{desert90}, which contains details on 
the \iras\ response functions in the different bands.
}
with the SED $L(\lambda)$ of the 
spectral type of each cluster galaxy, normalised according to 
the value of \lir\ (see Sect.~\ref{sec:model}). 

\begin{figure*}
\includegraphics[width=1.00\textwidth]{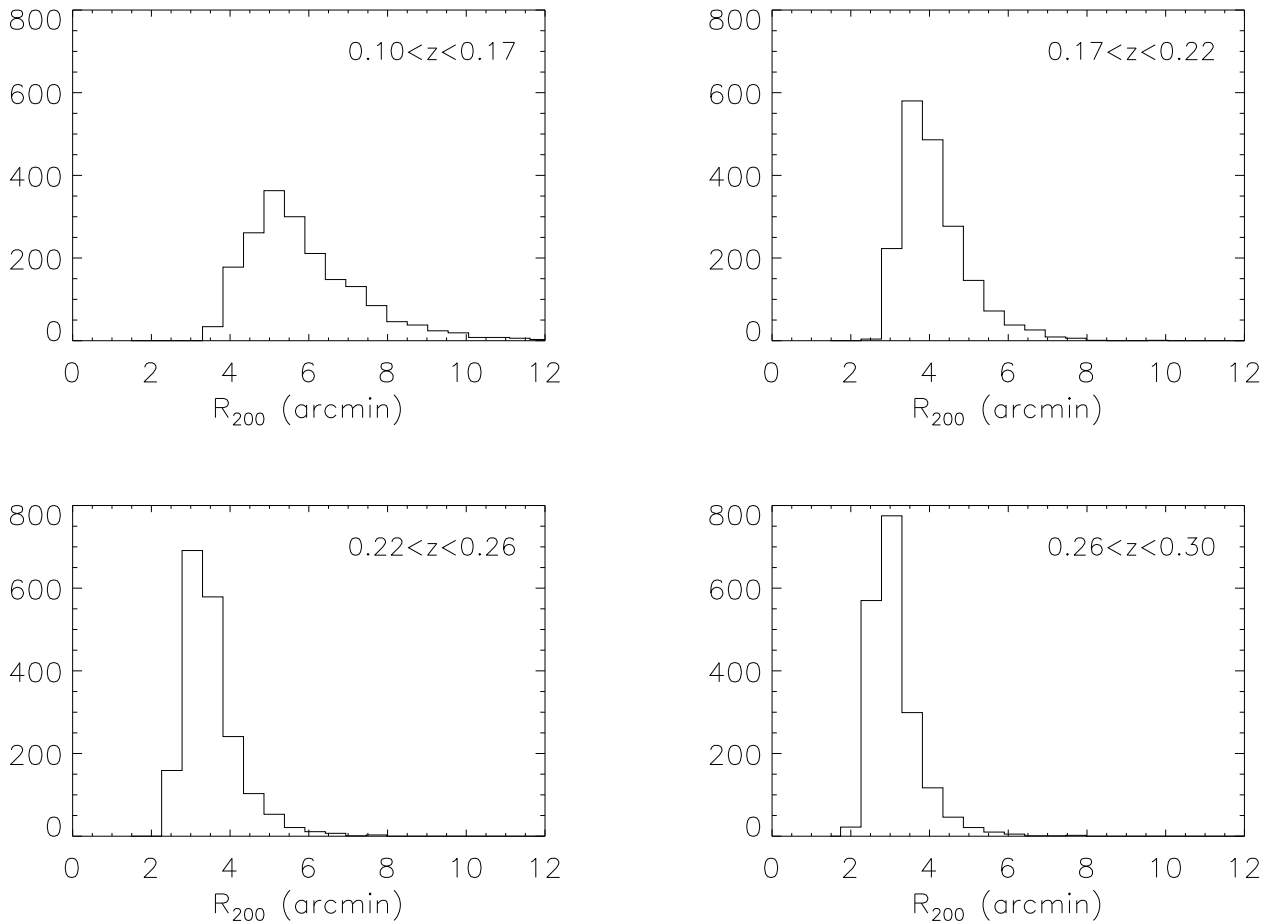}
\caption{Distribution of the angular sizes of $R_{200}$ for the cluster 
sample in 4 redshift intervals chosen in order to contain the same number of 
clusters (i.e. 1869). The bin size of each histogram is 0.5 arcmin.
}
\label{fig:hist_r200}
\end{figure*}

The stacked signals by \cite{giard08} are integrated
fluxes over an angular area of 10' of radius centered on the cluster
(see Sect.~\ref{sec:stack} for details). Since the \iras\ FWHM
beams are 4' and 4.5' at 60 and 100 \mum\, respectively, thus
comparable to the size of the observed field, we need to take into
account the possible loss of signal from galaxies
distant from the cluster center due to the convolution with the
instrumental beam. In order to model this effect, we
have to distribute the cluster galaxies in the
cluster potential well. We thus randomly assign to every galaxy a
distance from the BCG (considered to be at the cluster center, thus
at the center of the 10' field) by assuming that their spatial
distribution follows a NFW profile \citep{navarro97}. 
The two parameters needed
to characterise the NFW profile are $R_{200}$ and the concentration
$c$. They were obtained from the richness \ngal\ 
of each halo (see Sect.~\ref{ssec:maxbcg}) and by adopting the 
$N - M$ scaling relation 
of \cite{rykoff08} derived from the SDSS-maxBCG, and
the $(M,z)-c$ relation of \cite{dolag04} derived from numerical 
simulations. Then for every galaxy we compute the fraction of the 
signal that falls
inside the observed region in the two bands. This effect
proves to be completely negligible as globally less than one per 
cent of the signal falls outside the field in both bands.
Since galaxies of later spectral types are expected to be more 
spread than ellipticals, as a result of their recent accretion into 
the cluster  
\citep[see, e.g., the discussion in][]{popesso05}, 
we repeated our estimation on the possible 
lost signal by assuming that late-type galaxies are distributed 
uniformly in a sphere of radius $R_{200}$: even with this extreme 
hypothesis the amount of signal that is expected to fall outside the 
field remains negligible (about 3\% of the total).
The low impact of this effect on the global results 
can also be seen from the distributions of the 
angular sizes of $R_{200}$ obtained with this method shown in 
Fig.~\ref{fig:hist_r200} for different 
redshift bins. More than 70\% of the clusters have angular sizes 
lower than 5' and only in the lowest redshift bin some objects (43) 
exceed the 10' aperture.

With the stacking technique 
adopted by \cite{giard08}, it is clear that also the signal 
coming from foreground/background objects is present in the 
observed fields. Anyway, as said in Section~\ref{sec:stack}, 
the authors adopted a background subtraction based on the signal 
in the fields adjacent to every map, relying on the high statistical 
robustness of their sample. For this reason we can safely conclude 
that the emission of foreground/background objects has been 
successfully subtracted, thus we do not need to include it in 
our modelisation.


\section{Results and discussion}
\label{sec:results}

\subsection{Predicted and observed IR fluxes}

Our estimations of the 60 and 100~\mum\ fluxes from cluster 
galaxies are reported in  Table~\ref{tab:results}, 
associated to the emission calculated with our model 
for the different galaxy populations.

\begin{table}
\begin{center}
\caption{Estimated fluxes in the 60 \mum\ and 100 
  \mum\ \iras\ bands from the galaxy population of the SDSS-maxBCG
  clusters.
}

\begin{tabular}{lcrr}
\hline
\hline
Spectral type      & & $F_{60}$ & $F_{100}$ \\
                   & &  (Jy)    &   (Jy)    \\ 
\hline
\it{-Early-type}   & &          &           \\
BCG                & &     8.2  &    16.5   \\
E/S0               & &    31.1  &    62.9   \\
(Total early-type) & &   (39.3) &   (79.4)  \\
\\
\it{-Late-type}    & &          &           \\
Sa                 & &    36.4  &    58.6   \\
Sb                 & &   259.5  &   621.3   \\
Sc                 & &   333.9  &  1133.4   \\
Starburst          & &    15.4  &    12.1   \\
(Total late-type)  & &  (645.2) & (1825.4)  \\
\\
\hline
Total              & & 684.5$_{-138.0}^{+211.3}$  & 1904.8$_{-429.1}^{+617.1}$   \\
Observed           & & 570.1$\pm36.1$  & 1359.9$\pm249.1$  \\
\hline
\hline
\label{tab:results}
\end{tabular}
\end{center}
We show the total fluxes and contribution of each galaxy population 
included in our model. The quoted errors on the total values
correspond to the differences between our reference model and the
conservative and extreme scenarios described in Section~\ref{ssec:unc}.
The measurements by \cite{giard08} are reported together with their 
1$\sigma$ error bars.
\end{table}

The total flux due to the early-type galaxies estimated with our model
is 39.3~Jy at 60 \mum\ band and 79.4~Jy in the 100 \mum\ band,
accounting for about 7\% of and 6\%, respectively, of
the fluxes measured in the same bands by \cite{giard08}. In both bands
about 20\% of the E/S0 signal comes from the BCGs.

As expected, late-type galaxies constitute by far 
the most significant contribution to the IR emission. According to our 
reference model they contribute to 645.2~Jy at 60~\mum\ and 
1904.8~Jy at 100~\mum\ 
thus accounting for about 95\% of the total galactic 
emission. This contribution comes mostly from the Sb and Sc population, 
with the Sc being widely dominant in the 100~\mum\ band.

Despite their high luminosities, we predict that starburst galaxies do
not provide a significant contribution to the IR emission, due to
their low expected number. We obtain a contribution of
15.4~Jy at 60 \mum\ and 12.1~Jy at 100~\mum\,
corresponding respectively to about 2\% and 0.5\% of total predicted 
signal. Only in our extreme scenario their contribution becomes
non-negligible in the 60 \mum\ band, reaching the 5\% of
the total predicted flux. This low contribution agrees with the
low rate of starburst galaxies as found in the field by
\citet{lefloch05} at the redshift range of the SDSS-maxBCG
catalogue (i.e. $0.1<z<0.3$).

The total fluxes associated to the galactic emission predicted 
by our reference model are $684.5 \ [546.5,895.8]$~Jy at 60~\mum\ 
and $1904.8 \ [1475.7,2521.9]$~Jy at 100~\mum\ (the bracketed interval 
indicate the values derived from our conservative and extreme 
models, see Sect.~\ref{ssec:unc}). 
It appears that the reconstructed IR emission due to
the galactic dust emission of the cluster members can 
explain the entire signal measured by \cite{giard08}, with an 
indication that our reference model overestimates the total flux, 
particularly at 100~\mum. Indeed, these 
authors obtained $570.1\pm36.1$~Jy and $1359.9\pm249.1$~Jy at 
60 \mum\ and 100 \mum, respectively. We will propose an explanation 
of this discrepancy in the next sections. However, when considering 
our conservative scenario, the predicted emission is in good agreement  
with the total measured signal in both bands.

Given these results, modulo the uncertainties of our
modelisation, we obtain that the IR emission of the galaxy members 
is consistent with the total observed emission of our clusters 
sample, leaving little space to the possible presence of 
other components like intracluster dust.

\subsection{Redshift evolution}
\label{ssec:z_ev}

In this Section we characterise the redshift evolution of the average 
clusters luminosity. Following \cite{desert90}, we define the total 
luminosity in the 60 and 100 \mum\ \iras\ bands as
\begin{equation}
\label{eq:lum_60+100}
L_{60+100} \equiv 4 \pi d_{\rm L}^2 \times 
\left[ \lambda \frac{\Delta \nu}{\Delta \lambda} 
F_\lambda(60 \mu {\rm m}) + 
\lambda \frac{\Delta \nu}{\Delta \lambda}
F_\lambda(100 \mu {\rm m}) \right] \ ,
\end{equation}
where $\Delta \nu$ and $\Delta \lambda$ are the bandwidths in 
frequency and wavelength, respectively, of the two IRAS bands.
This formula represents a good approximation of the total luminosity of our 
haloes at $\lambda \gtrsim 40$\mum. 

\begin{figure}
\includegraphics[width=0.50\textwidth]{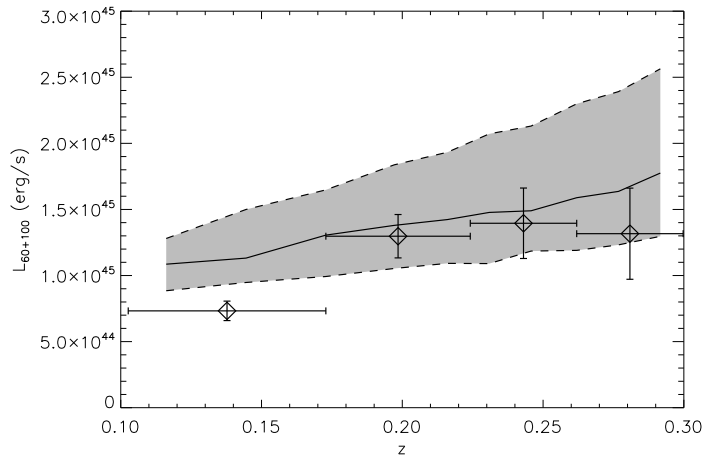}
\caption{Average cluster luminosities (see eq.~\ref{eq:lum_60+100}) 
  as a function of redshift.
  Diamonds represent the measurements and associated $1 \sigma$ error
  bars by \cite{giard08} in 4 redshift bins, defined in
  order to contains the same number of clusters (i.e. 1869). 
  The solid line represents the average luminosity of the clusters 
  in our reference model. The two dashed lines show the
  corresponding luminosity of the conservative and 
  extreme model (see Section~\ref{ssec:unc} for the details).}

\label{fig:lum_z}
\end{figure}

We show in Fig.~\ref{fig:lum_z} the average luminosity as obtained 
from eq.~\ref{eq:lum_60+100} for the clusters at different
redshifts, both as predicted by our model and as observed by
  \cite{giard08}. Our 4 bins are defined in order to contain the same
  number of clusters (i.e. 1869).
This analysis shows clearly that the discrepancy between 
our reference model with respect to the observed fluxes is due to the 
low-redshift clusters only, which are responsible for most 
of the predicted flux. In fact, while our reference model is compatible 
within $1 \sigma$ with the results of \cite{giard08} in the 3 
high-redshift bins, in the first bin ($0.10<z<0.17$)  the predicted 
luminosity is higher at $6 \sigma$ confidence. Our conservative 
scenario is also compatible with observed luminosities at $z>0.17$, 
while the discrepancy for the low-redshift clusters persists at 
more than $3 \sigma$. On the contrary, our extreme model is excluded 
by the results of \cite{giard08} through all the redshift range.

\subsection{Selection bias}

\begin{figure}
\includegraphics[width=0.50\textwidth]{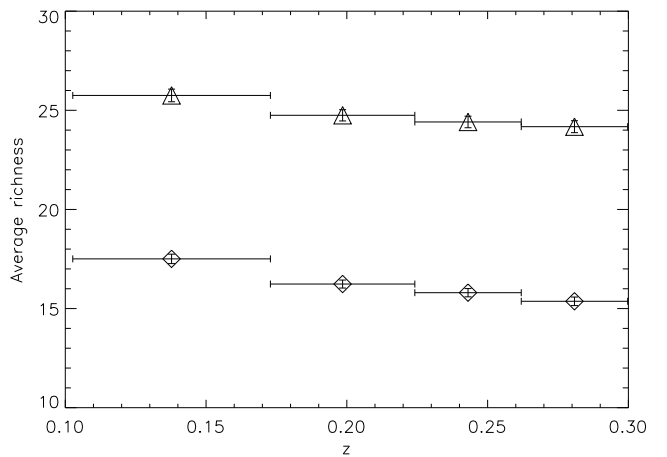}
\caption{Average richness as a function of 
redshift. Diamonds represent the average values of \ngal\ 
(i.e. the number of E/S0 members) with associated $1 \sigma$ 
errorbars in 4 redshift bins defined as in Fig.~\ref{fig:hist_r200} 
and \ref{fig:lum_z}.
Triangles represent the values of $N_{200}$ (i.e. total cluster 
members, see the definition in Section~\ref{ssec:late}).}
\label{fig:ngal_z}
\end{figure}

Our results indicate that although our model 
well describes the global IR emission of the SDSS-maxBCG 
galaxy clusters at $z \gtrsim 0.17$, the predicted galactic 
emission at low redshift clearly exceeds the measured stacked 
signal. The strong decrease in the IR luminosity is present also in the 
NED and NSC clusters samples analysed by \cite{giard08} and, as 
discussed in their work, it can be interpreted as a selection 
effect which biases towards rich and massive clusters at high 
redshift as 
it is evident for the NSC sample (see their Fig.~6). However, 
this is not the case for the clusters object of our analysis. In 
Fig.~\ref{fig:ngal_z} we show the average cluster richness in 
the same redshift bins both for the values of \ngal, 
directly obtained from the SDSS-maxBCG catalogue, and $N_{200}$ 
which includes late-type galaxies, as described in 
Section~\ref{ssec:late}. It is clear that no selection bias is 
present in our clusters sample: on the contrary low-redshift 
clusters show slightly higher values of \ngal. Even when including 
late-type galaxies this trend does not change significantly.

\subsection{Effect of the cluster environment}

These considerations indicate that the discrepancy 
between our model and the decrease of luminosity towards lower 
redshift must be connected with an evolution of the galaxy 
IR luminosity driven by the cluster environment, associated to gas 
and/or dust removal and consequent star-formation quenching. 
This can happen via ram-pressure stripping and tidal interaction. 
This picture is in agreement with the low number 
of high IR-to-optical galaxies observed in local galaxy clusters
\citep[see, e.g.,][]{bicay87}. A similar phenomenon 
is also seen in the observed properties of the galaxies of the Virgo 
cluster. In fact, \cite{boselli06b} observed a truncation in the 
disk of NGC 4569 both in the UV and in the IR (8--70 \mum) that they 
associate to a ram-pressure stripping of the external regions. 
More in general, \cite{gavazzi06} observe a consistent 
$H_\alpha$ deficiency in Virgo galaxies which indicates a low 
SFR and diminished dust heating/emission \citep[see also the 
discussion in][]{boselli06a}.
In this scenario, once the star-formation quenching happens, 
the SEDs of the galaxy members would be consistently modified 
at the IR and UV wavelengths, being not compatible anymore with 
field galaxies, thus explaining the excess predicted by our model. 
However, we must point out that the environmental effects on the Virgo 
cluster are likely to be much stronger than what is expected for 
the population of clusters considered in our analysis. In fact, 
the Virgo cluster is 
about one order of magnitude more massive than the majority of the 
SDSS-maxBCG galaxy clusters and it shows several evidences of recent 
dynamical activity. On the contrary, being the SDSS-maxBCG cluster 
sample mainly constituted by small haloes (see Section~\ref{ssec:maxbcg}), 
in the context of the 
hierarchical structure formation scenario it is reasonable to expect 
that they correspond to relaxed and dynamically old systems 
\citep[see, for instance, the results on the concentration-mass 
relation of][]{dolag04,buote07}.

\subsection{Constraints on intracluster dust}

If we take as a reference our conservative scenario and compare 
its results with the average observed luminosity in the 
three high-redshift bins, we can estimate an upper limit on the possible 
emission due to extragalactic dust of about 10\% of the total luminosity, 
which translates into a dust-to-gas mass abundance of 
$Z_{\rm d} \lesssim 5 \times 10^{-5}$ \citep[see the discussion 
in][]{giard08}. This figure is in agreement with the expectations from 
theoretical models \citep{popescu00,aguirre01} and with 
current observational upper limits on the dust abundance.
\cite{chelouche07} analysed the reddening of a sample of quasars 
in the direction of the clusters of the SDSS-maxBCG catalogue, 
obtaining an estimate of $Z_{\rm d} \approx 10^{-5}$. A similar 
result has also been obtained by \cite{bovy08} by measuring 
the dust absorption on the spectra of galaxies located behind local 
($z \sim 0.05$) galaxy clusters.

If on one side we do not see an evidence of intracluster dust 
emission in our sample, these last considerations leave an open 
question of how much dust has been lost by cluster galaxies polluting 
the ICM and if it can live enough to produce a significant diffuse 
IR emission. According to \cite{popescu00}, the dust stripped from 
infalling galaxies will probably remain localised close to their parent 
galaxies without diffusing efficiently into the ICM. 
In any case, the strong decrease in the total cluster luminosity 
observed by \cite{giard08} compared to our results, suggests that, 
if present, the signal of intracluster dust should be very small.


\section{Summary and conclusions}
\label{sec:concl}

In this work we performed, for the first time, a
thorough modelisation of the overall IR emission of galaxy clusters
due to cluster galaxies IR emission. We tested the results of our
model against the statistical stacking measurements of clusters IR
emission in the 60 and 100~\mum\ \iras\ bands by \cite{giard08},
making use of the SDSS-maxBCG catalogue of groups and clusters
\citep{koester07}.

We used the available SDSS-maxBCG data on the
luminosity in the $r$-band of the cluster early-type members, and
converted them into IR luminosities by adopting the model templates of
the \textsc{grasil} code \citep{silva98}.  Since the
SDSS-maxBCG catalogue does not contain information on the late-type
galaxies component, we used the results
on the morphology-density relation obtained by \cite{balogh04} on a
local SDSS galaxy sample to construct a model to
statistically associate to every cluster its spiral
galaxy population, namely the number of spiral members, their
spectral type and their $M_r$. Again, this 
information on the optical properties 
has been used to obtain the corresponding IR emission by making 
use of 4 other \textsc{grasil} templates to represent 
normal late-type (Sa, Sb and Sc) and starburst galaxies.
We also included in our model the expected redshift evolution of the
spiral fraction and the IR luminosity by using results on observed
galaxies in order to account for the possible difference in the spiral
galaxy population of the SDSS-maxBCG catalogue ($0.1<z<0.3$) and the more
local objects of the \cite{balogh04} sample ($z<0.08$).

Finally, we used our predictions to calculate the total expected 
galactic flux, by considering also the possible loss of signal 
due to the \iras\ beam smoothing and the \iras\ response 
function, and compared it with the measurements of \cite{giard08}. 

Our main results can be summarized as follows.

\begin{itemize}
\item[i)] According to our model, late-type galaxies
  represent about $\sim 35$\% of the galaxy population in nearby
  groups and clusters (i.e. $z<0.3$). This figure is coherent with the
  fact that this sample is dominated by groups and small clusters
  (\ngal$= 10-20$), thus letting this fraction of
    late-types falling between the field ($f_{\rm spi} \approx 0.5$)
  and the rich cluster ($f_{\rm spi} \approx 0.15$) regime.
\item[ii)] As expected, normal late-type galaxies constitute the most
  important contribution to the total IR luminosity
    at 60 and 100 \mum\, accounting for $\sim 95$\% of the
    total galactic emission. Oppositely, the impact of
  starburst galaxies is marginal.
\item[iii)] Early-type galaxies dominate the faint end of the LF 
            in the IR and they account for the remaining $\sim 5$\% 
            of the galactic contribution of the cluster IR emission.
\item[iv)] Our model shows that the total flux estimated from the galaxy 
           accounts, within uncertainties, for the entire stacked signal 
           measured by \cite{giard08}. With our reference model we 
           obtain $684.5$~Jy at 60 \mum\ and $1904.8$~Jy at 100~\mum, 
           which exceeds the stacked fluxes measured by \cite{giard08}. 
           However, when considering the uncertainties in our parameters, 
           we showed that in a conservative scenario this discrepancy 
           disappears.
\item[v)]  We compared the redshift distribution of the 
           IR emission predicted by our model to the measurements of 
           \cite{giard08} and found that the excess flux is present 
           only in the clusters at lower redshift ($z \lesssim 0.17$).
\end{itemize}

The results presented here show that the IR emission 
of galaxy clusters can be explained with its galactic component only, 
leaving very little room to the possible presence of any diffuse 
emission associated to intracluster dust. In this framework, the upper 
limit on the dust-to-gas mass abundance obtained by \cite{giard08} can 
be reduced by an order of magnitude, down to $Z_{\rm d} \lesssim 
5 \times 10^{-5}$. 
This result is in 
agreement with current estimations on the dust abundance 
in the ICM obtained from extinction and reddening measurements 
in the direction of SDSS clusters \citep{chelouche07,bovy08}.

The lack of diffuse dust, however, does not mean that the  
environment is not influencing the IR properties of its galaxy 
members. The fact that the predicted excess emission is concentrated 
in the lowest redshift objects indicates that the IR emission 
of late-type galaxies in local clusters is not completely 
compatible with their field equivalents. This result is in 
agreement with the lack of bright IR galaxies observed in several 
local clusters \citep[see][and references therein]{boselli06a} and
it is likely connected to the quenching of the star-forming 
activity driven by the cluster environment on its galaxy members, due 
to gas or dust removal: in this last case, the dust injected 
in the ICM is probably quickly depleted via sputtering processes.
However, it is not  clear why this could act significantly only 
at $z \lesssim 0.2$ when most of the large-scale structures are 
already formed. For this reason, and given the limited redshift 
range of the SDSS-maxBCG catalogue, it would be interesting to 
extend our analysis to higher redshifts, where the SFR is expected 
to be higher and more dynamical activity is expected at cluster 
scales. \\

\begin{acknowledgements}
We thank an anonymous referee who helped improving the presentation 
of our results. We are grateful to L. Silva for publishing some 
\textsc{grasil} templates in her webpage. We thank M. Balogh and T. 
F. Lagan\'a for 
providing some additional data not published in their papers.
We acknowledge useful discussions with S. Bardelli, J.-P. Bernard, 
A. Boselli, J. Lanoux, L. Pozzetti, F. Pozzi and C. Tonini. We are 
particularly grateful to A. Bongiorno 
for the help provided in the treatment of optical data. The authors 
L. Montier, E. Pointecouteau and M. Roncarelli acknowledge the support 
of grant ANR-06-JCJC-0141.
\end{acknowledgements}

\bibliographystyle{aa}

\newcommand{\noopsort}[1]{}

\label{lastpage}
\end{document}